\documentclass[
    ,final            
    ,sort&compress    
  ]
  {aipproc}

\layoutstyle{6x9}

\begin{document}
\newcommand{\chandra}{\textit{Chandra\/ }}
\newcommand{\lesssim}{{_<\atop^{\sim}}}
\newcommand{\msun}{$M_{\odot}$}
\newcommand{\mbh}{M_\bullet}
\newcommand{\mbul}{M_{\mathrm{bulge}}}
\newcommand{\hst}{\textit{HST\/}}
 
\newcommand{\apj}{ApJ}
\newcommand{\apjs}{ApJS}
\newcommand{\apjl}{ApJ}
\newcommand{\mnras}{MNRAS}
\newcommand{\pasp}{PASP}
\newcommand{\aap}{A\&A}
\newcommand{\aj}{AJ}
\newcommand{\araa}{ARA\&A}

\title{Finding Local Low-mass Supermassive Black Holes}


\author{Smita Mathur}{
  address={Department of Astronomy, The Ohio state University, 140
  W 18th Ave, Columbus, OH 43210, USA}  
}

\author{Himel Ghosh}{
  address={Department of Astronomy, The Ohio state University, 140
W 18th Ave, Columbus, OH 43210, USA} 
}

\author{Laura Ferrarese}{
  address={Herzberg Institute of Astrophysics, 5071 West Saanich Road, Victoria, BC V8X 4M6, Canada}
}

\author{Fabrizio Fiore}{
  address={INAF - Osservatorio Astronomico di Roma, via Frascati 33, 00040
Monteporzio Catone (Roma), Italy}
}

\begin{abstract}
The low-mass end of the supermassive black hole mass function is unknown
and difficult to determine. Here we discuss our successful program to
find active nuclei of late type ``normal'' galaxies using X-ray
detections and multiwavelength identifications. We conclude that most of
the \chandra detected nuclear X-ray sources are AGNs. We then outline
methods of black hole mass determination when broad emission lines are
unobservable.
\end{abstract}

\maketitle


\section{Introduction}

The local low-mass (below $10^6$ \msun) supermassive black hole (SMBH)
mass function is not well known. Current estimates of the SMBH mass
function are based on host galaxy properties (luminosity of the bulge
or bulge stellar velocity dispersion $\sigma$) and known scaling
relationships between the mass $\mbh$ of the SMBH and these properties
(most prominently $\mbh -
\sigma$ and $\mbh - \mbul$). These estimates can be widely discrepant
with each other at the low-mass end (see, for example, Fig.~8 in
\cite{gdea07}). 
Much of the uncertainty arises because it is unknown how the scaling
relationships extrapolate to very late-type and very low mass
galaxies. Yet SMBHs \emph{do} exist in very late-type spirals, e.g.\
NGC 4395, a spiral galaxy of type Sdm, with $\mbh \sim 3\times 10^5$
\msun\
\citep{pea05}, and in very low mass galaxies, e.g.\ POX 52, a dwarf galaxy,
with $\mbh \sim 3\times 10^5$ \msun\
\citep{bea04}.

Most of the measured SMBH masses are $\sim 10^8 $ \msun\ or greater,
however, as the sphere of influence of a less massive SMBH is
extremely hard to resolve even at moderate distances, even with the
Hubble Space Telescope (\hst). For example, the sphere of influence of
a $10^6 $ \msun\ SMBH at 15 Mpc is $\sim 30$ milliarcseconds. 
Since we cannot detect low-mass SMBHs by their dynamical
signature, looking for them by signs of their accretion activity may
be the only viable way of detecting them. An accreting low-mass SMBH
should be a low-luminosity active galactic nucleus (LLAGN).

How can an LLAGN be identified? Broad emission lines in the optical
spectrum allow an unambiguous confirmation of the presence of an
AGN. The sample of low-mass SMBHs in \citet{gh07a} was identified this
way. However, as the luminosity of an AGN decreases, the optical
spectrum of the galaxy nucleus becomes more and more dominated by host
galaxy light, and the signature of the AGN becomes difficult to
detect. For the lowest-luminosity AGNs, therefore, it is possible that
a system based on optical spectra would not classify the nuclei as
AGNs at all. Even when the optical spectrum shows no clear evidence of
an AGN, however, such evidence may still be present in other
wavelengths, such as x-ray and radio \citep{fea04}, and infrared
\citep{dea06,svea07,sea08}.

\section{Our Chandra program}

\subsection{Why X-ray selection?}

X-ray emission is an ubiquitous property of AGNs. If late type galaxies
indeed host supermassive black holes in their nuclei, and yet they have
not been discovered until now, they are most likely highly
obscured. This is less of a problem in X-rays than in optical
bands. X-ray observations with \chandra have uncovered AGNs in what were
thought to be normal galaxies in clusters \cite{mea02} and in fields
\cite{bea05}. X-ray surveys have also found new classes of AGNS,
e.g. ``X-ray bright optically normal galaxies''
\citep[XBONGS;][]{cea02}. Thus, it is highly likely that a strategy of
uncovering hidden AGNs in optically normal galaxies using X-ray
detections might work.

The sub-arcsecond angular resolution of \chandra is advantageous for
this program for identifying the true nucleus from surrounding galactic
sources. With these considerations we initiated a \chandra program to
search for active nuclei in centers of nearby late type galaxies
identified as X-ray sources. As we discuss below, this was relatively
easy; we detected nuclear X-ray sources in a large number of
galaxies. The second part, identifying the X-ray sources as accreting
SMBHs was far more difficult. We used a variety of techniques and
multiwavelength data for this purpose.

\subsubsection{The sample}

Our goal is to search for low-level nuclear activity in a representative
sample of low-mass galaxies within 20 Mpc. We selected 38 galaxies from
the NBG catalog \cite{t88} with following criteria: face-on spirals
(S0--Sdm) or dE and {\it NOT} known to host AGNs. We excluded starburst
galaxies from the sample to ease the identification of nuclear X-ray
sources. LINER galaxies were included because the nature of their energy
source is still a matter of debate.
Based on their optical luminosities ($-21< M_B
\le -15$) and morphology, most of these galaxies are expected to host
SMBHs with $10^5 \lesssim M_{\bullet} \lesssim 10^7$ \msun, with a few
objects hosting SMBHs as small as $10^4$\msun.

 With this study we hope to set firm lower limits to the number of
galaxies which host low mass BH, and to address the following
questions:\\
\noindent
-- What fraction of galaxies host
active SMBHs and what governs the activity of these SMBHs? \\
-- How does the fraction of active galaxies depend on the galaxy
morphology \& environment? \\
-- What is the primary component of a galaxy that relates to its BH
mass: the bulge or the dark matter halo? Finding active nuclei in
bulge-less spirals would imply the latter. \\
-- What is the relation between nuclear star formation activity and
 accretion onto a BH? \\
-- Is there a lower bound to the local BH mass function? \\
-- How do we understand BH growth and its
 relation to hierarchical galaxy formation?  
\cite[e.g.][]{mfpc04}

Of the 38 \chandra targets, 28 have been observed to
date. We also found 16 galaxies in the \chandra archive
meeting our selection criteria. Additionally, there were 6 galaxies in
the archive which met similar criteria applied to the RC3 catalog. We
discuss the latter sample first.



\section{Results}

\subsection{The archival sample I}

The six galaxies in this sample have morphological types Sa, Sb, Sc,
Scd (2) and Sd. All six have nuclear X-ray sources. For two galaxies,
NGC 4713 (Sd) and NGC 4647 (Sc), the data are inconclusive, but we
cannot rule out that they are AGNs. Interestingly, for two weak-bulge
galaxies NGC 5457 (Scd) and NGC 3184 (Scd) we can make a strong case
that they host AGNs; these galaxies are discussed in detail in Ghosh
et al. (these proceedings). The remaining two galaxies, NGC 3169 (Sa)
and NGC 4102 (Sb) are almost certainly AGNs; we discuss these here.

\begin{figure}
  \includegraphics[height=.3\textheight]{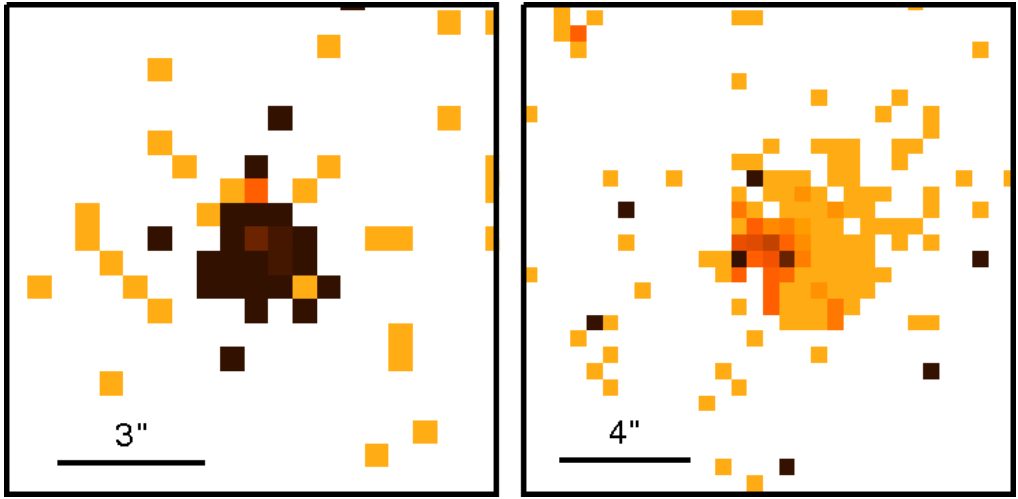}
  \caption{\chandra images of the nuclei of NGC 3169 (left) and NGC
  4102 (right). The pixels are colored according to hardness ratio $HR
  \equiv (H-S)/(H+S)$ where $H$ and $S$ are the counts in the 2.5--8 keV
  and 0.3--2.5 keV bands, respectively. The lightest pixels show the
  softest emission ($HR=-1$) and the darkest pixels the hardest
  emission ($HR=+1$). The bars in the lower left of each panel
  represent a projected distance of $\sim\! 300$ pc. North is up and
  east to the left in both panels.}
\label{fig:n31n41}
\end{figure}

{\bf NGC 3169:} This is an Sa galaxy at 20 Mpc, classified optically
as a low ionization nuclear emission line region (LINER). \chandra
detected a nuclear hard X-ray source (Fig.~\ref{fig:n31n41}),
indicative of an AGN. The X-ray spectrum is an absorbed power-law with
slope $\Gamma\sim 2$ and column density $N_H
\sim 10^{23}$ cm$^{-2}$. The AGN identification is clinched by the fact
that the X-ray source coincides with a 7mJy mas-scale VLA radio source
\cite{nfw05}. Thus the source is almost certainly an AGN.

{\bf NGC 4102:} This is a Sb galaxy at 17 Mpc, classified optically as
a HII nucleus. \chandra detected a hard point source in its nucleus
with extended soft emission (Fig.~\ref{fig:n31n41}). The spectrum of
the nuclear point source is a power-law with $\Gamma\sim 2$ and no
intrinsic absorption. However, a strong Fe K-$\alpha$ line is detected
with EW=2.5 keV. Even though the error on the EW is large, owing to
the low S/N of the spectrum, the EW is inconsistent with the ($\sim
300$ eV) value observed in unabsorbed AGNs. Thus the spectrum is that
of a reflection-dominated source. The
\chandra image and the spectrum are similar to those of Seyfert 2
galaxies \cite{gea07}. The nucleus is also a FIRST radio
source. Moreover, it is also an IR point source (2MASS) with luminosity
$10^{43}$ erg/s. This puts the source squarely in the AGN regime.

These observations show the power of X-ray detection and multiwavelength
identification of AGNs in what were thought to be ``normal''
galaxies. (See \cite{gmff08} for complete analysis of the sample.)

\subsection{The whole sample}

The six galaxies in the archival sample discussed above were scrutinized
one by for the possible signature of an AGN. This has been a hard job
owing to the low luminosity of the sources (except for the two best
cases discussed above). Dilution of the AGN signature due to
contamination from the host galaxy light then becomes a major
problem. Obscuration also affects the observed properties. As a result,
traditional diagnostics of finding AGNs, such as UV excess, X-ray to
optical flux ratio, X-ray to IR flux ratio, existence of broad emission
lines, line rations of narrow emission lines, all break down. Indeed, if
it were not a difficult task, we would have already known that these
galaxies host AGNs!

The identification problem becomes somewhat easier with a larger
sample. In addition to the six galaxies discussed above, we found 16
more from \citet{t88} in the Chandra archive following our selection
criteria. Nuclear X-ray sources were detected in 12. Thus, 18 of the 22
archival galaxies host a nuclear X-ray source. Moreover, in most cases,
the nucleus was the brightest X-ray source, in one the second brightest
X-ray source and in one the only X-ray source. It is highly unlikely
that the brightest X-ray binary, a star, or any other normal galactic
X-ray source resides in the nucleus most of the time. Thus we conclude
that most of the \chandra detected nuclear X-ray sources are AGNs.

Our new \chandra survey was much shallower than the archival
observations. Preliminary analysis indicates that 6 out of 28 galaxies
host a nuclear X-ray source.

\section{Discussion}

\subsection{How to measure BH masses}

Reverberation mapping of broad emission lines provides a standard method
for measuring BH masses in AGNs. Alternatively, scaling relations using
the width of broad emission lines and AGN luminosity are used. How
should we measure BH masses in our newly discovered AGNs which do not
show broad emission lines? This can be done in a variety of ways:

\noindent
1. As noted above, host galaxy contamination is a major contributor
toward hiding the AGN signature of our low-luminosity sources. Broad
emission lines of these AGNs might have been swamped by the host galaxy
light in ground-based spectra. Therefore it is imperative to obtain
spectra of these sources with the highest possible spatial resolution from
space, e.g. with STIS on HST. This might uncover broad emission lines.

\noindent
2. Construction of power-density spectra (PDS) using X-ray variability
analysis provides a powerful tool for BH mass determination
\cite{mpea04}. The break frequency in PDS is found to correlate
strongly with the BH mass; since our sources are X-ray detected, this
is potentially a highly feasible method.

\noindent
3. The X-ray power-law slope correlates strongly with the Eddington
luminosity ration \cite{wmp04}. Measuring the X-ray slope and
luminosity would then estimate the BH mass.

\noindent
4. SMBH mass is known to correlate with the bulge velocity dispersion
\cite{fm00,gea00} and the bulge luminosity \cite{kr95,md01,mh03}. These relations can be
used for BH mass determination, once we ascertain the existence of a
SMBH. This method, however, will not work for bulge-less
galaxies. Moreover, if we do not have independent measurements of BH
masses, we cannot extend these known correlations to low BH masses.

\noindent
5. SMBH mass is also known to correlate with the circular velocity of
the host galaxy \cite{f02,bea03}. This method will have the same
caveat mentioned above, though it works for bulge-less galaxies as
well.

Thus there are several ways to estimate masses of SMBHs which we uncover
using X-ray detections. This is obviously a hard job; at this point,
however, we are at the step of finding these low-mass SMBHs in the first
place.

\subsection{What's Next?}

In order to make progress in finding and identifying low-mass SMBHs and
measuring their masses several steps need to be taken.

\noindent
1. Through our \chandra survey and the accompanying archival programs we
have realized that traditional diagnostics for identifying AGNs fail for
low-luminosity AGNs embedded in luminous galaxies. We need to develop
new diagnostic tools using multi-wavelength data for this purpose.

\noindent
2. Deep X-ray spectra of likely AGNs. This will help secure AGN
identification. We have obtained XMM-{\it Newton} data on one of our
candidate.

\noindent
3. HST STIS spectroscopy. As noted above, this will uncover broad
emission lines if any and possibly change the narrow emission line flux
ratios.

\noindent
4. HST ACS imaging. This will help obtain true colors of the AGN, after
removing a substantial contribution of the host galaxy. We can them
identify an AGN with traditional ``UV excess'' and/or with broad-band
spectral energy distribution.

\noindent
5. IR diagnostics: {\it Spitzer} IRAC colors. While AGNs are blue in the
optical, they are red in mid-IR. AGNs occupy a distinct location in the IRAC
color-color plot, allowing identification of X-ray sources \cite{skea05}.

\noindent
6. IR diagnostics: near IR spectra. High ionization IR emission lines
with IRS on {\it Spitzer} have been used to uncover AGNs by
\citet{sea08}. With the end of {\it Spitzer} cryogenic mission, IRS
is no longer available. Finding near-IR ``coronal'' lines might prove
useful for the same purpose.

\noindent
7. Radio observations. Finding a radio point source coincident with the
X-ray point source will help secure the identification. VLA observations
of our candidate AGNs is an obvious next step.

\noindent
8. Bigger \chandra/{\it Spitzer} survey will provide better statistics
on fraction of galaxies hosting AGNs. A deeper survey will allow us to
determine AGN duty cycle to lower luminosities.

\noindent
9. Hard X-ray observations. A survey of nearby galaxies in hard X-rays
will detect and identify nuclear SMBHs (as well as off-nuclear accreting
BHs). The Indian X-ray mission ASTROSAT, which has hard X-ray
capabilities is due to launch in 2009. ASTROSAT observations will be
invaluable in our BH quest.


\begin{theacknowledgments}
This work is supported in part by the National Aeronautics and Space
Administration through Chandra Award Number GO7-8111X issued by the
Chandra X-ray Observatory Center, which is operated by the Smithsonian
Astrophysical Observatory for and on behalf of the National Aeronautics
Space Administration under contract NAS8-03060.

\end{theacknowledgments}



\bibliographystyle{aipproc}   


\begin{thebibliography}{25}
\expandafter\ifx\csname natexlab\endcsname\relax\def\natexlab#1{#1}\fi
\providecommand{\enquote}[1]{``#1''}
\expandafter\ifx\csname url\endcsname\relax
  \def\url#1{\texttt{#1}}\fi
\expandafter\ifx\csname urlprefix\endcsname\relax\def\urlprefix{URL }\fi
\providecommand{\eprint}[2][]{\url{#2}}

\bibitem[{Graham} et~al.(2007)]{gdea07}
A.~W. {Graham}, S.~P. {Driver}, P.~D. {Allen}, and J.~{Liske}, \emph{\mnras}
  \textbf{378}, 198--210 (2007), \eprint{arXiv:0704.0316}.

\bibitem[{Peterson} et~al.(2005)]{pea05}
B.~M. {Peterson}, M.~C. {Bentz}, L.-B. {Desroches}, A.~V. {Filippenko}, L.~C.
  {Ho}, S.~{Kaspi}, A.~{Laor}, D.~{Maoz}, E.~C. {Moran}, R.~W. {Pogge}, and
  A.~C. {Quillen}, \emph{\apj} \textbf{632}, 799--808 (2005),
  \eprint{arXiv:astro-ph/0506665}.

\bibitem[{Barth} et~al.(2004)]{bea04}
A.~J. {Barth}, L.~C. {Ho}, R.~E. {Rutledge}, and W.~L.~W. {Sargent},
  \emph{\apj} \textbf{607}, 90--102 (2004), \eprint{arXiv:astro-ph/0402110}.

\bibitem[{Greene} and {Ho}(2007)]{gh07a}
J.~E. {Greene}, and L.~C. {Ho}, \emph{\apj} \textbf{656}, 84--92 (2007),
  \eprint{arXiv:astro-ph/0608061}.

\bibitem[{Filho} et~al.(2004)]{fea04}
M.~E. {Filho}, F.~{Fraternali}, S.~{Markoff}, N.~M. {Nagar}, P.~D. {Barthel},
  L.~C. {Ho}, and F.~{Yuan}, \emph{\aap} \textbf{418}, 429--443 (2004),
  \eprint{arXiv:astro-ph/0401593}.

\bibitem[{Dale} et~al.(2006)]{dea06}
D.~A. {Dale}, J.~D.~T. {Smith}, L.~{Armus}, B.~A. {Buckalew}, G.~{Helou}, R.~C.
  {Kennicutt}, Jr., J.~{Moustakas}, H.~{Roussel}, K.~{Sheth}, G.~J. {Bendo},
  D.~{Calzetti}, B.~T. {Draine}, C.~W. {Engelbracht}, K.~D. {Gordon}, D.~J.
  {Hollenbach}, T.~H. {Jarrett}, L.~J. {Kewley}, C.~{Leitherer}, A.~{Li},
  S.~{Malhotra}, E.~J. {Murphy}, and F.~{Walter}, \emph{\apj} \textbf{646},
  161--173 (2006), \eprint{arXiv:astro-ph/0604007}.

\bibitem[{Satyapal} et~al.(2007)]{svea07}
S.~{Satyapal}, D.~{Vega}, T.~{Heckman}, B.~{O'Halloran}, and R.~{Dudik},
  \emph{\apjl} \textbf{663}, L9--L12 (2007), \eprint{arXiv:0706.1050}.

\bibitem[{Satyapal} et~al.(2008)]{sea08}
S.~{Satyapal}, D.~{Vega}, R.~P. {Dudik}, N.~P. {Abel}, and T.~{Heckman},
  \emph{\apj} \textbf{677}, 926--942 (2008), \eprint{arXiv:0801.2759}.

\bibitem[{Martini} et~al.(2002)]{mea02}
P.~{Martini}, D.~D. {Kelson}, J.~S. {Mulchaey}, and S.~C. {Trager},
  \emph{\apjl} \textbf{576}, L109--L112 (2002),
  \eprint{arXiv:astro-ph/0208017}.

\bibitem[{Brand} et~al.(2005)]{bea05}
K.~{Brand}, A.~{Dey}, M.~J.~I. {Brown}, C.~R. {Watson}, B.~T. {Jannuzi}, J.~R.
  {Najita}, C.~S. {Kochanek}, J.~C. {Shields}, G.~G. {Fazio}, W.~R. {Forman},
  P.~J. {Green}, C.~J. {Jones}, A.~T. {Kenter}, B.~R. {McNamara}, S.~S.
  {Murray}, M.~{Rieke}, and A.~{Vikhlinin}, \emph{\apj} \textbf{626}, 723--732
  (2005), \eprint{arXiv:astro-ph/0503156}.

\bibitem[{Comastri} et~al.(2002)]{cea02}
A.~{Comastri}, M.~{Mignoli}, P.~{Ciliegi}, P.~{Severgnini}, R.~{Maiolino},
  M.~{Brusa}, F.~{Fiore}, A.~{Baldi}, S.~{Molendi}, R.~{Morganti},
  C.~{Vignali}, F.~{La Franca}, G.~{Matt}, and G.~C. {Perola}, \emph{\apj}
  \textbf{571}, 771--778 (2002), \eprint{arXiv:astro-ph/0202080}.

\bibitem[{Tully}(1988)]{t88}
R.~B. {Tully}, \emph{{Nearby galaxies catalog}}, Cambridge and New York,
  Cambridge University Press, 1988, 221 p., 1988.

\bibitem[{Menci} et~al.(2004)]{mfpc04}
N.~{Menci}, F.~{Fiore}, G.~C. {Perola}, and A.~{Cavaliere}, \emph{\apj}
  \textbf{606}, 58--66 (2004), \eprint{arXiv:astro-ph/0401261}.

\bibitem[{Nagar} et~al.(2005)]{nfw05}
N.~M. {Nagar}, H.~{Falcke}, and A.~S. {Wilson}, \emph{\aap} \textbf{435},
  521--543 (2005), \eprint{astro-ph/0502551}.

\bibitem[{Ghosh} et~al.(2007)]{gea07}
H.~{Ghosh}, R.~W. {Pogge}, S.~{Mathur}, P.~{Martini}, and J.~C. {Shields},
  \emph{\apj} \textbf{656}, 105--115 (2007), \eprint{arXiv:astro-ph/0606461}.

\bibitem[{Ghosh} et~al.(2008)]{gmff08}
H.~{Ghosh}, S.~{Mathur}, F.~{Fiore}, and L.~{Ferrarese}, \emph{\apj},
  submitted, \eprint{arXiv:astro-ph/0801.4382}.

\bibitem[{McHardy} et~al.(2004)]{mpea04}
I.~M. {McHardy}, I.~E. {Papadakis}, P.~{Uttley}, M.~J. {Page}, and K.~O.
  {Mason}, \emph{\mnras} \textbf{348}, 783--801 (2004),
  \eprint{arXiv:astro-ph/0311220}.

\bibitem[{Williams} et~al.(2004)]{wmp04}
R.~J. {Williams}, S.~{Mathur}, and R.~W. {Pogge}, \emph{\apj} \textbf{610},
  737--744 (2004), \eprint{arXiv:astro-ph/0402236}.

\bibitem[{Ferrarese} and {Merritt}(2000)]{fm00}
L.~{Ferrarese}, and D.~{Merritt}, \emph{\apjl} \textbf{539}, L9--L12 (2000),
  \eprint{arXiv:astro-ph/0006053}.

\bibitem[{Gebhardt} et~al.(2000)]{gea00}
K.~{Gebhardt}, R.~{Bender}, G.~{Bower}, A.~{Dressler}, S.~M. {Faber}, A.~V.
  {Filippenko}, R.~{Green}, C.~{Grillmair}, L.~C. {Ho}, J.~{Kormendy}, T.~R.
  {Lauer}, J.~{Magorrian}, J.~{Pinkney}, D.~{Richstone}, and S.~{Tremaine},
  \emph{\apjl} \textbf{539}, L13--L16 (2000), \eprint{arXiv:astro-ph/0006289}.

\bibitem[{Kormendy} and {Richstone}(1995)]{kr95}
J.~{Kormendy}, and D.~{Richstone}, \emph{\araa} \textbf{33}, 581--624 (1995).

\bibitem[{McLure} and {Dunlop}(2001)]{md01}
R.~J. {McLure}, and J.~S. {Dunlop}, \emph{\mnras} \textbf{327}, 199--207
  (2001), \eprint{arXiv:astro-ph/0009406}.

\bibitem[{Marconi} and {Hunt}(2003)]{mh03}
A.~{Marconi}, and L.~K. {Hunt}, \emph{\apjl} \textbf{589}, L21--L24 (2003),
  \eprint{arXiv:astro-ph/0304274}.

\bibitem[{Ferrarese}(2002)]{f02}
L.~{Ferrarese}, \emph{\apj} \textbf{578}, 90--97 (2002),
  \eprint{arXiv:astro-ph/0203469}.

\bibitem[{Baes} et~al.(2003)]{bea03}
M.~{Baes}, P.~{Buyle}, G.~K.~T. {Hau}, and H.~{Dejonghe}, \emph{\mnras}
  \textbf{341}, L44--L48 (2003), \eprint{arXiv:astro-ph/0303628}.

\bibitem[{Stern} et~al.(2005)]{skea05}
D.~{Stern}, P.~{Eisenhardt}, V.~{Gorjian}, C.~S. {Kochanek}, N.~{Caldwell},
  D.~{Eisenstein}, M.~{Brodwin}, M.~J.~I. {Brown}, R.~{Cool}, A.~{Dey},
  P.~{Green}, B.~T. {Jannuzi}, S.~S. {Murray}, M.~A. {Pahre}, and S.~P.
  {Willner}, \emph{\apj} \textbf{631}, 163--168 (2005),
  \eprint{arXiv:astro-ph/0410523}.

\end{thebibliography}


\end{document}